\documentclass[aps,prd,onecolumn,groupedaddress,showpacs,nofootinbib,amssymb]{revtex4}
\usepackage[dvips]{graphicx}
\usepackage{amssymb}
\usepackage{amsmath}
\usepackage{graphicx,,color}
\usepackage{amsfonts}
\usepackage{bm}
\usepackage{cancel}
\usepackage{comment}

\newcommand\be{\begin{equation}}
\newcommand\ee{\end{equation}}

\allowdisplaybreaks[4]

\begin{document}

\tolerance=5000

\title{Einstein-Gauss-Bonnet Cosmological Theories at Reheating and at the End of the Inflationary Era}
\author{V.K. Oikonomou$^{1,2}$}
\email{Corresponding author:
voikonomou@gapps.auth.gr;v.k.oikonomou1979@gmail.com}
\author{Pyotr Tsyba$^{2}$}\,
\email{pyotrtsyba@gmail.com}
\author{Olga Razina$^{2}$}\,
\email{olvikraz@mail.ru} \affiliation{$^{1)}$Department of
Physics, Aristotle University of Thessaloniki, Thessaloniki 54124,
Greece} \affiliation{$^{2)}$L.N. Gumilyov Eurasian National
University - Astana, 010008, Kazakhstan}

 \tolerance=5000

\begin{abstract}
In this work we study the GW170817-compatible
Einstein-Gauss-Bonnet theories during the reheating and the end of
inflationary era. Given the scalar field potential $V(\phi)$ which
can have some intrinsic importance for the theory, determining the
scalar coupling function $\xi(\phi)$ can be cumbersome due to lack
of analyticity. The GW170817 observation constrains the scalar
coupling function and the scalar field potential to have some
interdependence, thus during the slow-roll era one can calculate
the scalar coupling function. However, when the slow-roll era
ends, it is expected that the scalar coupling function should have
a different form and the same applies for the reheating era,
assuming that the scalar potential of the theory does not change.
In this work we exactly aim to highlight this feature of
Einstein-Gauss-Bonnet theories, as the Universe evolves through
distinct sequential evolution eras, and we focus on how to
determine the scalar coupling function during the various
evolutionary eras, from inflation to the reheating era. Regarding
both the end of the inflationary era and the reheating era, it is
found that the Hubble rate obeys a constant-roll-like condition of
the form $\dot{H}=\delta H^2$, thus the determination of the
scalar Gauss-Bonnet function $\xi(\phi)$ is reduced to solving a
differential equation. A mentionable feature of the era exactly at
the end of inflation is that the Klein-Gordon equation is
decoupled from the field equations, because the Gauss-Bonnet
invariant is zero. We provide several examples of interest to
support our arguments.
\end{abstract}


\maketitle

\section{Introduction}

Undoubtedly, during the next decade inflationary theories
\cite{inflation1,inflation2,inflation3,inflation4} will be in the
mainstream of interest since they will be directly tested either
by stage four Cosmic Microwave Background experiments
\cite{CMB-S4:2016ple,SimonsObservatory:2019qwx}, or indirectly via
gravitational wave experiments
\cite{Hild:2010id,Baker:2019nia,Smith:2019wny,Crowder:2005nr,Smith:2016jqs,Seto:2001qf,Kawamura:2020pcg,Bull:2018lat,LISACosmologyWorkingGroup:2022jok}
which will capture the stochastic imprint of inflation. Among the
candidate theories, Einstein-Gauss-Bonnet theories
\cite{Hwang:2005hb,Nojiri:2006je,Cognola:2006sp,Nojiri:2005vv,Nojiri:2005jg,Satoh:2007gn,Bamba:2014zoa,Yi:2018gse,Guo:2009uk,Guo:2010jr,Jiang:2013gza,Kanti:2015pda,vandeBruck:2017voa,Kanti:1998jd,Pozdeeva:2020apf,Vernov:2021hxo,Pozdeeva:2021iwc,Fomin:2020hfh,DeLaurentis:2015fea,Chervon:2019sey,Nozari:2017rta,Odintsov:2018zhw,Kawai:1998ab,Yi:2018dhl,vandeBruck:2016xvt,Kleihaus:2019rbg,Bakopoulos:2019tvc,Maeda:2011zn,Bakopoulos:2020dfg,Ai:2020peo,Easther:1996yd,Antoniadis:1993jc,Antoniadis:1990uu,Kanti:1995vq,Kanti:1997br,Odintsov:2020mkz,Odintsov:2020zkl,Odintsov:2023aaw}
serve as an important class of theories, which in some cases
predict a blue-tilted tensor spectrum and also are theoretically
motivated since these contain string corrections. The striking
GW170817 observation back in 2017 altered our perception for
several candidate inflationary theories, including the
Einstein-Gauss-Bonnet theories. The reason was the fact that the
kilonova GW170817 event indicated that the gravitational wave
speed is almost equal to light speed. Einstein-Gauss-Bonnet
theories, and other similar string corrected theories, predict a
propagation speed for tensor perturbations that are not equal to
that of light, thus these classes of theories were put into
existential peril, see
\cite{Ezquiaga:2017ekz,Baker:2017hug,Creminelli:2017sry,Sakstein:2017xjx}
for a discussion on which theories contradict GW170817. However,
if one requires that the propagation speed of tensor perturbations
is equal to that of light, the theoretical issues of
Einstein-Gauss-Bonnet gravity may be refined, and this was done in
a series of articles
\cite{Oikonomou:2021kql,Oikonomou:2022xoq,Odintsov:2020sqy}. In
these theories, the Gauss-Bonnet scalar coupling function $\xi
(\phi)$ satisfies a differential equation $\ddot{\xi}=H\dot{\xi}$
and in effect the scalar potential and the scalar Gauss-Bonnet
coupling are connected and are not free to choose. The formalism
developed in
\cite{Oikonomou:2021kql,Oikonomou:2022xoq,Odintsov:2020sqy} leads
to GW170817-compatible and viable inflationary theories. However,
given the scalar potential, the scalar Gauss-Bonnet coupling
function can be found only semi-analytically, and only for the
evolutionary patch of the Universe which generates a specific
Hubble rate. Thus in principle, one cannot know the exact form of
the scalar Gauss-Bonnet coupling function, and only approximate
forms of it may be found. In this article we aim to highlight this
important feature in Einstein-Gauss-Bonnet theories of gravity,
focusing on two eras that follow the inflationary era, the end of
inflation era and the reheating era. In both these cases, the
Hubble rate and its time derivative satisfy a constant-roll-like
relation $\dot{H}=\delta H^2$, thus this constrains and simplifies
the field equations. In effect, the scalar Gauss-Bonnet function
obeys a set of differential equations and may be obtained
analytically for each evolutionary era, given the scalar
potential. We use several illustrative examples, and we
demonstrate that indeed the scalar Gauss-Bonnet coupling acquires
distinct functional forms during the various distinct evolutionary
eras of our Universe. Also for a specific viable inflationary
case, we examine the behavior of the Gauss-Bonnet term after
inflation and during the inflationary era, and we show that for
this interesting example, the effect of the Gauss-Bonnet coupling
is minor, thus the reheating era is mainly controlled by the
scalar potential.

\section{A Brief Overview of GW170817-compatible Einstein-Gauss-Bonnet Framework}

Let us start our analysis by recalling how the Gauss-Bonnet scalar
coupling function is obtained during the inflationary era, by
using the conditions required so that the gravitational wave speed
is equal to unity in natural units, and also the slow-roll
conditions. We consider the Einstein-Gauss-Bonnet gravitational
action,
\begin{equation}
\label{action} \centering
S=\int{d^4x\sqrt{-g}\left(\frac{R}{2\kappa^2}-\frac{1}{2}\partial_{\mu}\phi\partial^{\mu}\phi-V(\phi)-\frac{1}{2}\xi(\phi)\mathcal{G}\right)}\,
,
\end{equation}
with $R$ standing for the Ricci scalar, $\kappa=\frac{1}{M_p}$
with being $M_p$ the reduced Planck mass, and also $\mathcal{G}$
denotes the Gauss-Bonnet invariant in four dimensions, which is
$\mathcal{G}=R^2-4R_{\alpha\beta}R^{\alpha\beta}+R_{\alpha\beta\gamma\delta}R^{\alpha\beta\gamma\delta}$
where $R_{\alpha\beta}$ and $R_{\alpha\beta\gamma\delta}$ denote
the Ricci and Riemann tensors respectively. For the rest of the
article we shall assume that the geometric background is a flat
Friedmann-Lemaitre-Robertson-Walker (FLRW) metric, with line
element,
\begin{equation}
\label{metric} \centering
ds^2=-dt^2+a(t)^2\sum_{i=1}^{3}{(dx^{i})^2}\, ,
\end{equation}
where $a(t)$ denotes as usual the scale factor, and for the FLRW
background, the Gauss-Bonnet invariant becomes
$\mathcal{G}=24H^2(\dot H+H^2)$, with $H$ being the Hubble rate
$H=\frac{\dot{a}}{a}$. Notice that one requires that an
inflationary era occurs, one must have $\dot{H}\ll H^2$, however
when inflation ends, at this time instance, we have that the
Gauss-Bonnet invariant is exactly zero, since $\dot{H}=-H^2$. We
shall consider this scenario later on in this article. Now,
assuming that the scalar field has only a time-dependence, upon
varying the gravitational action (\ref{action}) with respect to
the metric and the scalar field, we obtain the field equations,
\begin{equation}
\label{motion1} \centering
\frac{3H^2}{\kappa^2}=\frac{1}{2}\dot\phi^2+V+12 \dot\xi H^3\, ,
\end{equation}
\begin{equation}
\label{motion2} \centering \frac{2\dot
H}{\kappa^2}=-\dot\phi^2+4\ddot\xi H^2+8\dot\xi H\dot H-4\dot\xi
H^3\, ,
\end{equation}
\begin{equation}
\label{motion3} \centering \ddot\phi+3H\dot\phi+V'+12 \xi'H^2(\dot
H+H^2)=0\, .
\end{equation}
The inflationary era is realized when $\dot{H}\ll H^2$ and also we
assume that the slow-roll conditions hold true for the scalar
field, so in total we have the following conditions holding true
during the inflationary era,
\begin{equation}\label{slowrollhubble}
\dot{H}\ll H^2,\,\,\ \frac{\dot\phi^2}{2} \ll V,\,\,\,\ddot\phi\ll
3 H\dot\phi\, .
\end{equation}
Now the GW170817 event imposed serious constraints on the
gravitational wave speed of inflationary theories. The propagation
speed of the tensor perturbations of a flat FLRW metric has the
form,
\begin{equation}
\label{GW} \centering c_T^2=1-\frac{Q_f}{2Q_t}\, ,
\end{equation}
with the functions $Q_f$, $F$ and $Q_b$ defined above for
Einstein-Gauss-Bonnet theories being equal to $Q_f=8
(\ddot\xi-H\dot\xi)$, $Q_t=F+\frac{Q_b}{2}$,
$F=\frac{1}{\kappa^2}$ and $Q_b=-8 \dot\xi H$. Hence in order to
have exactly $c_T^2=1$, we must have $Q_f=0$, which constrains the
Gauss-Bonnet scalar coupling function to obey the differential
equation $\ddot\xi=H\dot\xi$. We can express this differential
equation in terms of the scalar field as follows,
\begin{equation}
\label{constraint1} \centering
\xi''\dot\phi^2+\xi'\ddot\phi=H\xi'\dot\phi\, ,
\end{equation}
where the ``prime'' denotes differentiation with respect to the
scalar field, and also we used $\dot\xi=\xi'\dot\phi$ and
$\frac{d}{dt}=\dot\phi\frac{d}{d\phi}$. By making the assumption,
\begin{equation}\label{firstslowroll}
 \xi'\ddot\phi \ll\xi''\dot\phi^2\, ,
\end{equation}
which is actually motivated by the slow-roll conditions obeyed by
the scalar field, Eq. (\ref{constraint1}) becomes,
\begin{equation}
\label{constraint} \centering
\dot{\phi}\simeq\frac{H\xi'}{\xi''}\, .
\end{equation}
Then Eqs. (\ref{motion3}) and (\ref{constraint}) yield,
\begin{equation}
\label{motion4} \centering \frac{\xi'}{\xi''}\simeq-\frac{1}{3
H^2}\left(V'+12 \xi'H^4\right)\, .
\end{equation}
Clearly this is a differential equation that relates the scalar
field potential $V(\phi)$ with the scalar Gauss-Bonnet function
$\xi (\phi)$, used in the previous literature
\cite{Oikonomou:2021kql,Oikonomou:2022xoq,Odintsov:2020sqy}. This
can be further simplified, by assuming that,
\begin{equation}\label{mainnewassumption}
\kappa \frac{\xi '}{\xi''}\ll 1\, ,
\end{equation}
and furthermore the following assumptions,
\begin{equation}\label{scalarfieldslowrollextra}
12 \dot\xi H^3=12 \frac{\xi'^2H^4}{\xi''}\ll V\, ,
\end{equation}
which are related to the constraint of Eq.
(\ref{mainnewassumption}). By combining Eqs.
(\ref{slowrollhubble}), (\ref{constraint}) and
(\ref{scalarfieldslowrollextra}), we get the simplified form of
the field equations,
\begin{equation}
\label{motion5} \centering H^2\simeq\frac{\kappa^2V}{3}\, ,
\end{equation}
\begin{equation}
\label{motion6} \centering \dot H\simeq-\frac{1}{2}\kappa^2
\dot\phi^2\, ,
\end{equation}
\begin{equation}
\label{motion8} \centering \dot\phi\simeq\frac{H\xi'}{\xi''}\, .
\end{equation}
Also, due to Eq. (\ref{motion5}), the condition
(\ref{scalarfieldslowrollextra}) acquires the simpler form,
\begin{equation}\label{mainconstraint2}
 \frac{4\kappa^4\xi'^2V}{3\xi''}\ll 1\, .
\end{equation}
Furthermore, the differential equation (\ref{motion4}), takes the
simpler form,
\begin{equation}
\label{maindiffeqnnew} \centering
\frac{V'}{V^2}+\frac{4\kappa^4}{3}\xi'\simeq 0\, ,
\end{equation}
and constrains basically the scalar potential and the scalar
Gauss-Bonnet coupling during the inflationary era. It is very
important to further stress that the differential equation
(\ref{maindiffeqnnew}) holds true only during the inflationary era
and also by assuming a slow-rolling scalar field, so this is not a
general relation between the scalar field potential and the scalar
Gauss-Bonnet coupling function. It is the aim of this work to find
the relation of the scalar Gauss-Bonnet coupling and the scalar
potential, at the end of inflation and during the reheating era.
Note that, this difficulty arises due to the difficulty to solve
the field equations analytically, given the potential, and perhaps
the Hubble rate. But even if these last two were given, one would
still have to determine the scalar Gauss-Bonnet coupling function
during the various stages of evolution, for example during the
inflationary era for a quasi-de Sitter evolution, during the
reheating and so on, with only the differential equation
$\ddot{\xi}=H\dot{\xi}$ holding true. The lack of analyticity
forces us to break the problem to smaller problems, and thus this
shows the motivation of this work. Now, having the differential
equation (\ref{maindiffeqnnew}) at hand, we can determine the
scalar potential and find the scalar Gauss-Bonnet coupling
function. We shall consider three potentials that may lead to
analytic results in the case of reheating era and the end of
inflation era, two of which we present here. Consider the
following scalar potential,
\begin{equation} \label{potA} \centering
V(\phi)= \beta  \exp (\kappa  \lambda  \phi )\, ,
\end{equation}
where $\beta$ is a constant with mass dimensions $[M]^4$. By
solving the differential equation (\ref{maindiffeqnnew}), we
obtain the following scalar Gauss-Bonnet function,
\begin{equation}
\label{modelA} \xi(\phi)=\frac{3 e^{-\kappa  \lambda  \phi }}{4
\beta  \kappa ^4}\, .
\end{equation}
Also consider the following scalar potential,
\begin{equation} \label{potA1} \centering
V(\phi)=   \lambda  \kappa \phi^n\, ,
\end{equation}
where $\lambda$ in this case is a constant with mass dimensions
$[M]^{5-n}$. By solving the differential equation
(\ref{maindiffeqnnew}), we obtain the following scalar
Gauss-Bonnet function,
\begin{equation}
\label{modelA1} \xi(\phi)=\frac{3 \phi ^{-n}}{4 \kappa ^5 \lambda
}\, .
\end{equation}
We shall compare these solutions for the Gauss-Bonnet scalar
coupling function $\xi(\phi)$ to the solutions that will be
obtained from the end of inflation era and from the reheating era.

\section{Einstein-Gauss-Bonnet Gravity During the Reheating Era and at the End of Inflation}

Having discussed the inflationary era of the Einstein-Gauss-Bonnet
gravity, let us now consider the reheating era and the end of
inflation era. In the reheating era, the scale factor of the
Universe is $a(t)\sim t^{1/2}$, thus, the Hubble rate and its time
derivative satisfy a constant-roll like relation of the form
$\dot{H}=\beta H^2$, and specifically,
\begin{equation}\label{constantroll1}
\dot{H}=-2 H^2\, ,
\end{equation}
so by taking this into account and also the fact that
$\ddot{\xi}=H\dot{\xi}$, the field equations (\ref{motion1}) and
(\ref{motion2}) yield the following constraint for the scalar
field,
\begin{equation}\label{scalarfieldconstraint1}
\dot{\phi}^2=2V\, .
\end{equation}
Now, in ordinary scalar field theory in the absence of
Gauss-Bonnet corrections, the field equation (\ref{motion3})
combined with the constraint (\ref{scalarfieldconstraint1}) would
yield an exponential potential for the scalar field. However, due
to the presence of the Gauss-Bonnet corrections, the constraint
(\ref{scalarfieldconstraint1}) can be inserted and used in the
field equation for the scalar field, thus one obtains the
following differential equation for the Gauss-Bonnet scalar
coupling function,
\begin{equation}\label{diffeqn1scalarcoupling}
2V'+3H\dot{\phi}-12\xi'H^4=0\, .
\end{equation}
From the Friedmann equation, by using the constraint
(\ref{scalarfieldconstraint1}), we get,
\begin{equation}\label{constraintscalarderivativeh}
H\dot{\phi}==\frac{\frac{3H^2}{\kappa^2}-2V}{12\xi'H^2}\, ,
\end{equation}
thus combining Eq. (\ref{diffeqn1scalarcoupling}) and the
Friedmann equation we get,
\begin{equation}\label{hubblealgebraic}
\frac{3H^3}{\kappa^2}-7HV=-V'\sqrt{2V}\, ,
\end{equation}
where we used that $\dot{\phi}=-\sqrt{2V}$ because the scalar
field is expected to decrease as the time evolves. The algebraic
equation (\ref{hubblealgebraic}) can be solved with respect to the
Hubble rate and yields the following real solution,
\begin{align}\label{hubblealgrebaricsolution}
& H(V,V')=\frac{7 \sqrt[3]{2} \kappa ^2 V(\phi )}{3
\sqrt[3]{\sqrt{2} \sqrt{81 \kappa ^4 V(\phi ) V'(\phi )^2-686
\kappa ^6 V(\phi )^3}-9 \sqrt{2} \kappa ^2 \sqrt{V(\phi )} V'(\phi
)}}\\ \notag & +\frac{\sqrt[3]{\sqrt{2} \sqrt{81 \kappa ^4 V(\phi
) V'(\phi )^2-686 \kappa ^6 V(\phi )^3}-9 \sqrt{2} \kappa ^2
\sqrt{V(\phi )} V'(\phi )}}{3 \sqrt[3]{2}}\, .
\end{align}
Using Eqs. (\ref{hubblealgrebaricsolution}),
(\ref{diffeqn1scalarcoupling}) and also the fact that
$\dot{\phi}=-\sqrt{2V}$, one obtains a differential equation for
the Gauss-Bonnet scalar coupling function $\xi (\phi)$, which is
the following,
\begin{equation}\label{finalreheatingdifferentialequation}
2V'-3H(V,V')\sqrt{2V}-12\xi'H(V,V')^4=0\, ,
\end{equation}
so given the scalar potential, the above can yield the function
$\xi (\phi)$ during the reheating era by solving the differential
equation, or simply find $\xi'(\phi)$ by solving the algebraic
equation (\ref{finalreheatingdifferentialequation}) with respect
to $\xi'(\phi)$. Now let us investigate the solutions for the
function $\xi(\phi)$, for the potentials used in the previous
section, and we compare the new solutions for $\xi(\phi)$ during
the reheating era and the inflationary era. Let us start with the
exponential potential of Eq. (\ref{potA}) in which case the scalar
Gauss-Bonnet function for a slow-roll inflationary era yields the
solution (\ref{modelA}). If we plug the scalar potential
(\ref{potA}) in Eq. (\ref{hubblealgrebaricsolution}) we get,
\begin{equation}\label{hubblescalar1}
H(\phi)=A e^{\frac{\kappa  \lambda  \phi }{2}}\, ,
\end{equation}
where $A$ stands for,
\begin{equation}\label{parameterA}
A=\frac{2^{2/3} \left(\sqrt{\beta ^3 \kappa ^6 \left(81 \lambda
^2-686\right)}-9 \beta ^{3/2} \kappa ^3 \lambda \right)^{2/3}+14
\beta  \kappa ^2}{3\ 2^{5/6} \sqrt[3]{\sqrt{\beta ^3 \kappa ^6
\left(81 \lambda ^2-686\right)}-9 \beta ^{3/2} \kappa ^3 \lambda
}}\, .
\end{equation}
Using the solution (\ref{hubblescalar1}) and plugging it in the
differential equation (\ref{finalreheatingdifferentialequation})
we get,
\begin{equation}\label{finalexpsolution}
\xi(\phi)=\frac{e^{-\kappa  \lambda  \phi } \left(3 \sqrt{2} A
\sqrt{\beta }-2 \beta  \kappa  \lambda \right)}{12 A^4 \kappa
\lambda }\, .
\end{equation}
Apparently, this is a similar solution to that of Eq.
(\ref{modelA}) but this is an accidental feature because of the
simplicity of the model. Thus in this case, the scalar coupling
function during the inflationary era is a rescaled function of the
reheating era scalar coupling function.
\begin{figure}
\centering
\includegraphics[width=25pc]{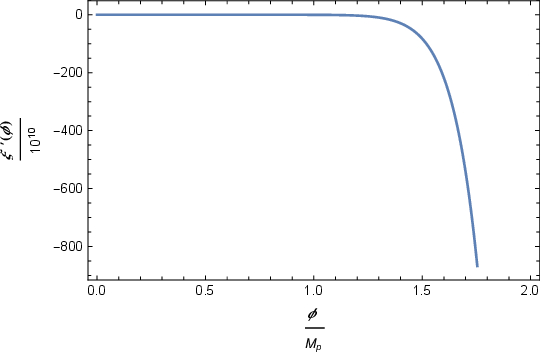}
\caption{The behavior of the scalar Gauss-Bonnet coupling function
$\xi'(\phi)$ during the reheating era for the scalar potential
(\ref{potA1viable}). }\label{plot1}
\end{figure}
This feature is model dependent though, as we now show. Let us
assume that the scalar potential has the form (\ref{potA1}), and
in the inflationary era the scalar Gauss-Bonnet coupling function
is given by relation (\ref{modelA1}). Now during the reheating
era, by plugging the potential (\ref{potA1}) in Eq.
(\ref{hubblealgrebaricsolution}) we get,
\begin{equation}\label{hubblescalar12}
H(\phi)=\frac{14 \kappa ^3 \lambda  \phi ^n+2^{2/3} S(\phi
)^{2/3}}{3\ 2^{5/6} \sqrt[3]{S(\phi )}}\, ,
\end{equation}
where the function $S(\phi)$ is defined as follows,
\begin{equation}\label{sphi}
S(\phi)= \frac{\phi  \sqrt{\kappa ^7 \lambda ^3 \phi ^{3 n-2}
\left(81 n^2-686 \kappa ^2 \phi ^2\right)}-9 \kappa ^2 n
\left(\kappa  \lambda  \phi ^n\right)^{3/2}}{\phi }\, .
\end{equation}
Using the solution (\ref{hubblescalar12}) and plugging it in the
differential equation (\ref{finalreheatingdifferentialequation})
we get,
\begin{equation}\label{finalexpsolution2}
\xi'(\phi)=\frac{a_1(\phi ) S(\phi )^{2/3}+a_2(\phi )
\sqrt[3]{S(\phi )}+a_5(\phi )}{a_3(\phi ) S(\phi )^{2/3}+a_4(\phi
) \sqrt[3]{S(\phi )}+a_6(\phi )}\, ,
\end{equation}
where the functions $Q(\phi)$, $a_1(\phi)$, $a_2(\phi)$,
$a_3(\phi)$, $a_4(\phi)$, $a_5(\phi)$ and $a_6(\phi)$ are defined
as follows,
\begin{align}\label{definitionsai}
& Q(\phi)= \sqrt{-\kappa ^7 \lambda ^3 \phi ^{3 n-2} \left(686
\kappa ^2 \phi ^2-81 n^2\right)}\, ,
\\ \notag & a_1(\phi)= 9\ 2^{2/3} \sqrt{\kappa } \sqrt{\lambda } \phi ^{2-\frac{n}{2}} Q(\phi )-81\ 2^{2/3} \kappa ^4 \lambda ^2 n \phi ^{n+1}\, ,
\\ \notag & a_2(\phi)= 162 \sqrt[3]{2} \kappa ^{9/2} \lambda ^{5/2} n^2 \phi ^{\frac{3 n}{2}}-18 \sqrt[3]{2} \kappa  \lambda  n \phi  Q(\phi )\, ,
\\ \notag & a_3(\phi)= -108\ 2^{2/3} \kappa ^7 \lambda ^3 n^2 \phi ^{2 n}-1372\ 2^{2/3} \kappa ^9 \lambda ^3 \phi ^{2 n+2}+12\ 2^{2/3} \kappa ^{7/2} \lambda ^{3/2} n \phi ^{\frac{n}{2}+1} Q(\phi )\, ,
\\ \notag & a_4(\phi)= 3528 \sqrt[3]{2} \kappa ^9 \lambda ^3 n \phi ^{2 n+1}-392 \sqrt[3]{2} \kappa ^6 \lambda ^2 \phi ^{n+2} Q(\phi )\, ,
\\ \notag & a_5(\phi)= 126 \kappa ^{7/2} \lambda ^{3/2} \phi ^{\frac{n}{2}+2} Q(\phi )-1134 \kappa ^7 \lambda ^3 n \phi ^{2 n+1}\, ,
\\ \notag & a_6(\phi)= -6048 \kappa ^{10} \lambda ^4 n^2 \phi ^{3 n}+19208 \kappa ^{12} \lambda ^4 \phi ^{3 n+2}+672 \kappa ^{13/2} \lambda ^{5/2} n \phi ^{\frac{3 n}{2}+1} Q(\phi
)\, .
\end{align}
Hence it is obvious that the scalar Gauss-Bonnet function during
the reheating and the slow-roll era for the scalar potential
(\ref{potA1}) are quite distinct. The two models we presented are
just demonstrational potentials and will not be further studied.
However, an interesting question is the following: suppose we
started with a specific potential and the corresponding
inflationary scalar Gauss-Bonnet function which both generate a
viable inflationary era, like for example the following two which
were developed in Ref.
\cite{Oikonomou:2021kql,Oikonomou:2022xoq,Odintsov:2020sqy},
\begin{equation} \label{potA1viable} \centering
V(\phi)=  \frac{3}{4 \beta  \kappa ^{\nu +4} \phi ^{\nu }+3 \gamma
\kappa ^4}\, ,
\end{equation}
where $\beta$ and $\gamma$ are two dimensionless constants. By
solving the differential equation (\ref{maindiffeqnnew}), we
obtain the following scalar Gauss-Bonnet function,
\begin{equation}
\label{modelA1viable} \xi(\phi)=\beta  \kappa ^{\nu } \phi ^{\nu
}\, ,
\end{equation}
what is the behavior of the Gauss-Bonnet scalar coupling function
during the reheating era? This will indicate whether the
Gauss-Bonnet string corrections affect the reheating era, at least
in this case. This can easily be examined by using the potential
(\ref{potA1viable}) and plugging it in Eq.
(\ref{hubblealgrebaricsolution}) and we get,
\begin{equation}\label{hubblescalar12viable}
H(\phi)=\frac{7 \kappa ^2 Q(\phi )+\left(18 \sqrt{2} \beta  \nu
\kappa ^{\nu +6} \phi ^{\nu -1} Q(\phi )^{5/2}+S(\phi
)\right)^{2/3}}{\sqrt{3} \sqrt[3]{18 \sqrt{2} \beta  \nu  \kappa
^{\nu +6} \phi ^{\nu -1} Q(\phi )^{5/2}+S(\phi )}}\, ,
\end{equation}
where,
\begin{align}\label{definitionsaiviable}
& S(\phi)=\sqrt{\frac{648 \beta ^2 \nu ^2 \kappa ^{2 \nu +6} \phi
^{2 \nu -2}-343 \kappa ^8 \left(4 \beta  \kappa ^{\nu } \phi ^{\nu
}+3 \gamma \right)^2}{\kappa ^{14} \left(4 \beta  \kappa ^{\nu }
\phi ^{\nu }+3 \gamma \right)^5}}\, ,
\\ \notag &
Q(\phi)= \frac{1}{\kappa ^4 \left(4 \beta  \kappa ^{\nu } \phi
^{\nu }+3 \gamma \right)}\, .
\end{align}
Using the solution (\ref{hubblescalar12viable}) and plugging it in
the differential equation
(\ref{finalreheatingdifferentialequation}) we get,
\begin{equation}\label{finalexpsolution2}
\xi'(\phi)=-\frac{3 R(\phi ) \left(18 \sqrt{2} \beta  \nu  \kappa
^{\nu +6} \phi ^{\nu -1} Q(\phi )^{5/2}+S(\phi )\right)^{4/3}}{4
\left(7 \kappa ^2 Q(\phi )+\left(18 \sqrt{2} \beta  \nu  \kappa
^{\nu +6} \phi ^{\nu -1} Q(\phi )^{5/2}+S(\phi
)\right)^{2/3}\right)^4}\, ,
\end{equation}
where $R(\phi)$ is defined to be equal to,
\begin{equation}\label{rphi}
R(\phi)=24 \beta  \nu  \kappa ^{\nu +4} \phi ^{\nu -1} Q(\phi
)^2+\frac{3 \sqrt{2} \sqrt{Q(\phi )} \left(7 \kappa ^2 Q(\phi
)+\left(18 \sqrt{2} \beta  \nu  \kappa ^{\nu +6} \phi ^{\nu -1}
Q(\phi )^{5/2}+S(\phi )\right)^{2/3}\right)}{\sqrt[3]{18 \sqrt{2}
\beta  \nu  \kappa ^{\nu +6} \phi ^{\nu -1} Q(\phi )^{5/2}+S(\phi
)}}\, .
\end{equation}
Apparently, the scalar Gauss-Bonnet function during the reheating
and the slow-roll era for the scalar potential (\ref{potA1viable})
are quite distinct, so let us see how the scalar Gauss-Bonnet
function behaves during the reheating era.
\begin{figure}
\centering
\includegraphics[width=25pc]{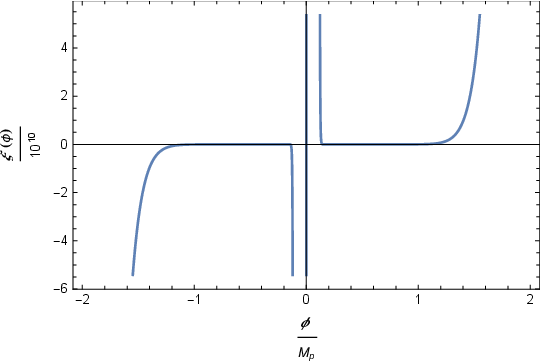}
\caption{The behavior of the scalar Gauss-Bonnet coupling function
$\xi(\phi)$ at exactly the end of the inflationary era for the
scalar potential (\ref{potA1viable}). }\label{plot2}
\end{figure}
A set of the parameter values that yield a viable inflationary era
are $(\beta,\gamma,\nu)=(1.2\times 10^{10},1000,15)$, and for
these values in Fig. \ref{plot1} we plot $\xi'(\phi)$ as a
function of the scalar field. As time evolves the scalar field
values decrease, thus the reheating era is approached as the
scalar field values tend to zero. As it can be seen, the
derivative of the scalar coupling function tends to zero, and
therefore this means that $\xi(\phi)\sim \mathrm{const}$.
Therefore, the Gauss-Bonnet term in the Lagrangian is effectively
$\sim \mathcal{G}$ thus the action integral of this term is just a
surface term, and therefore for this model, the Gauss-Bonnet
correction terms play no role in the reheating era, and therefore
the reheating era is solely determined by the behavior of the
scalar field potential.

Having discussed the behavior of the scalar Gauss-Bonnet coupling
function during the reheating era, now we shall consider the end
of inflation era, that is, the exact time instance that inflation
ends. In this case, the total effective equation of state
parameter is $w=-\frac{1}{3}$ thus we have $\dot{H}=-H^2$ which is
again a constant-roll behavior. In ordinary scalar field theory,
this would result to an exponential potential for the scalar
field, but in the Einstein-Gauss-Bonnet case, this constraints the
scalar Gauss-Bonnet function as we now evince. By taking into
account the fact that $\dot{H}=-H^2$ and also the fact that
$\ddot{\xi}=H\dot{\xi}$, the field equations (\ref{motion1}) and
(\ref{motion2}) yield the following constraint for the scalar
field,
\begin{equation}\label{scalarfieldconstraint1end}
\dot{\phi}^2=V\, .
\end{equation}
As we mentioned, in ordinary scalar field theory in the absence of
Gauss-Bonnet corrections, the field equation (\ref{motion3})
combined with the constraint (\ref{scalarfieldconstraint1end})
would yield an exponential potential for the scalar field but in
our case, it will simply constrain the scalar Gauss-Bonnet
function. From Eq. (\ref{scalarfieldconstraint1end}) we have
$\ddot{\phi}=\frac{V'}{2}$. Now, due to the fact that
$\dot{H}=-H^2$, the scalar field equation of motion becomes,
\begin{equation}\label{motion3decoupled}
\ddot{\phi}+3H\dot{\phi}+V'=0\, ,
\end{equation}
so by substituting $\ddot{\phi}=\frac{V'}{2}$ and Eq.
(\ref{scalarfieldconstraint1end}) and taking the square, we have,
\begin{equation}\label{anotherintreeqn}
H^2=\left(\frac{V'}{2V} \right)^2\, ,
\end{equation}
so by substituting this in the Friedmann equation (\ref{motion1})
we finally have,
\begin{equation}\label{finalenddiffequation}
\frac{3}{\kappa^2}\left(\frac{V'}{2V}
\right)^2=\frac{3V}{2}-6\xi'\left(\frac{V'}{2V} \right)^2V'\, ,
\end{equation}
which is a differential equation that, given the scalar potential,
can be solved with respect to $\xi (\phi)$, or it can be viewed as
an algebraic equation that can yield $\xi'(\phi)$. Now let us
consider the examples we gave in the case of the reheating era,
and we compare the solutions for the scalar Gauss-Bonnet
couplings. Consider first the scalar potential of Eq.
(\ref{potA}), so plugging this exponential potential in the
differential equation (\ref{finalenddiffequation}), we easily get
the following solution for $\xi(\phi)$,
\begin{equation}\label{xiphi1}
\xi(\phi)=\frac{\frac{\lambda  e^{-\kappa  \lambda  \phi }}{\beta
\kappa }+2 \phi }{2 \kappa ^3 \lambda ^3}\, ,
\end{equation}
which is quite different from the expression of Eq.
(\ref{modelA}). Accordingly, for the potential of Eq.
(\ref{potA1}), the scalar Gauss-Bonnet coupling function at the
end of the inflationary era is,
\begin{equation}\label{xiphi12}
\xi(\phi)=\frac{\phi ^4}{4 n^3}-\frac{\phi ^{2-n}}{2 \kappa ^3
\lambda  (2-n) n}\, ,
\end{equation}
which is quite distinct from that of Eq. (\ref{modelA1}). Finally,
for the potential of Eq. (\ref{potA1viable}) which recall that it
generates a viable inflationary era, the solution of the
differential equation (\ref{finalenddiffequation}) is,
\begin{align}\label{xiphi123}
& \xi(\phi)=\frac{\phi ^2 \kappa ^{-3 \nu } \left(\frac{128 \beta
^4 \nu ^2 \kappa ^{4 \nu +2} \phi ^{\nu }}{\nu +2}+\frac{162 \beta
\gamma ^2 \kappa ^{\nu } \phi ^{2-2 \nu }}{\nu -2}+\frac{81 \gamma
^3 \phi ^{2-3 \nu }}{3 \nu -4}\right)}{192 \beta ^3 \nu ^3} \\
\notag &  +\frac{\phi ^2}{192 \beta ^3 \nu ^3} \kappa ^{-3 \nu }
\left(-\frac{72 \beta ^2 \gamma  \kappa ^{2 \nu } \phi ^{-\nu }
\left(\gamma  \kappa ^2 (\nu -4) \nu ^2-6 (\nu -2) \phi
^2\right)}{(\nu -4) (\nu -2)}-48 \beta ^3 \kappa ^{3 \nu }
\left(\phi ^2-2 \gamma  \kappa ^2 \nu ^2\right)\right) \, ,
\end{align}
which is of course quite different from that of Eq.
(\ref{modelA1viable}). Now for this last solution it is worth to
investigate the behavior of the scalar coupling function as a
function of the scalar field. To this end, for the same set of
parameter values used in the reheating era case, that is, for
$(\beta,\gamma,\nu)=(1.2\times 10^{10},1000,15)$, which recall
that these yield a viable inflationary era, in Fig. \ref{plot2} we
plot $\xi(\phi)$ as a function of the scalar field. In this case
too, as time evolves, the scalar field values decrease, hence we
focus on values of the scalar field in the sub-Planck region. As
it can be seen, the scalar Gauss-Bonnet coupling function is
nearly zero for sub-Planck values of the scalar field, but for
nearly zero values, the scalar Gauss-Bonnet coupling function
blows up. However, it is not certain that the scalar field
actually took such values at the end of inflation, so considering
that the scalar field values at the end of inflation are of the
order of the Planck mass, this means that the scalar Gauss-Bonnet
coupling function is nearly zero, thus it does not affect the
evolution at the end of inflation. This is an interesting feature,
combined with the one we obtained in the reheating era case
developed previously.

In conclusion, in this section we demonstrated that given the
scalar potential, it is actually difficult to have an analytic
form of the scalar coupling Gauss-Bonnet function, if this is
assumed that it realizes the various evolutionary patches of our
Universe. Thus we presented the formalism on how to find the
analytic form of $\xi (\phi)$ or its derivative $\xi (\phi)$,
during the inflationary era, the reheating era and the end of
inflation era. This may yield important information for the
reheating era, and to which extent it affects the reheating and
subsequent radiation domination era. This formalism can also be
applied in other modified gravities, so we defer this task for
future works.

\section*{Conclusions}

In this work we aimed to demonstrate that the scalar Gauss-Bonnet
coupling function can be found analytically in distinct
evolutionary epochs of the Universe, and that in general it has
distinct approximate forms for these epochs. We considered
Einstein-Gauss-Bonnet gravities, which are compatible with the
GW170817 event. We demonstrated that during the inflationary era,
the scalar Gauss-Bonnet coupling and the scalar potential are
connected and satisfy a specific differential equation, which
yields the approximate form of the scalar Gauss-Bonnet coupling
during the slow-roll era. During the end of inflation era, the
Hubble rate and its time derivative satisfy a constant-roll
evolution of the form $\dot{H}=-H^2$ and thus the scalar potential
and the scalar Gauss-Bonnet coupling satisfy a different
differential equation which provides us with a different solution
for the scalar Gauss-Bonnet coupling. Furthermore, during the
reheating era, the Hubble rate and its time derivative satisfy an
alternative constant-roll evolution of the form $\dot{H}=-2H^2$
and therefore the scalar potential and the scalar Gauss-Bonnet
coupling satisfy another distinct from the other two differential
equation, which provides us with a different solution for the
scalar Gauss-Bonnet coupling. We presented several illustrative
and easy to study analytically examples, and we showed explicitly
that the scalar Gauss-Bonnet coupling indeed is different during
the various evolutionary epochs we considered. We also presented a
model which generates a viable inflationary era, and we showed
that the Gauss-Bonnet coupling tends to zero at the end of
inflation and during the reheating era. Thus for this example, the
Gauss-Bonnet term has a minimal effect on the reheating era, and
therefore the latter is affected solely by the scalar potential.
The formalism we presented in this paper can also be used in other
modified gravity frameworks and we aim to study such scenarios in
future works. Finally, the inflationary framework of theories at
hand can produce a blue-tilted tensor spectrum, and this could be
valuable in view of the latest NANOGrav observation in 2023
\cite{NANOGrav:2023gor}, although an extremely high blue-tilted
tensor spectral index would be needed \cite{Vagnozzi:2023lwo}. One
question would be, if the reheating era can impose some
constraints on the functional form of the Gauss-Bonnet coupling
during inflation, and in effect on the gravitational waves,
however no tensor perturbations are generated during the reheating
and at the end of inflation. Hence these eras are disconnected and
thus unrelated.

\section*{Acknowledgments}

This research has been is funded by the Committee of Science of
the Ministry of Education and Science of the Republic of
Kazakhstan (Grant No. AP19674478).

\end{document}